\documentclass[multphys,vecphys]{svmult}

\usepackage{makeidx}     
\usepackage{graphicx}    
\usepackage{multicol}    

\makeindex             

\begin{document}

\title*{Low frequency radio and X-ray properties of core-collapse supernovae}
\titlerunning{Radio and X-ray properties of supernovae}
\author{A. Ray\inst{1}\and
P. Chandra\inst{1,2}\and F. Sutaria\inst{3}\and S. Bhatnagar\inst{4,5}}
\authorrunning{A. Ray, P. Chandra, F. Sutaria, S. Bhatnagar} 
\institute{Tata Institute of Fundamental Research, Mumbai, India
\texttt{akr@tifr.res.in}
\and Joint Astronomy Program, Indian Inst. Science, Bangalore
\texttt{poonam@tifr.res.in}
\and Physics Dept., Technical University Munich, Germany
\texttt{fsutaria@ph.tum.de}
\and National Centre for Radio Astrophysics, TIFR, Pune, India
\and National Radio Astronomy Observatory, Socorro, USA
\texttt{sbhatnag@aoc.nrao.edu}
}
%
%
\maketitle

\begin{abstract}
 Radio and X-ray studies of young supernovae probe the
interaction between the supernova shock waves and the
surrounding medium and give clues to the nature and past
of the progenitor star. Here we discuss 
the early emission from type Ic SN 2002ap and argue that
repeated Compton boosting of optical photons
by hot electrons presents the most natural explanation of the prompt
X-ray emission. We describe the radio spectrum of another
type Ic SN 2003dh (GRB030329) obtained with combined GMRT and
VLA data. We report on the low frequency radio monitoring of SN 1995N
and our objectives of 
distinguishing between competing models
of X-ray emission from this SN and the nature of
its progenitor by X-ray spectroscopy. 
Radio studies on SN 2001gd, SN 2001ig
and SN 2002hh are mentioned.
\end{abstract}

\section{Introduction}
\label{sec:introduction}

The association of long duration Gamma Ray Bursts with core
collapse supernovae is by now established with
SN 2003dh (GRB030329) \cite{Hjo03}.
Observational evidence supports the viewpoint that GRBs 
are energetic explosions like supernovae occuring in
star formation regions and a large fraction of their energy
is directed in relativisitic jets.
Supernovae are known to be explosions of massive (and intermediate
mass) stars, although the nature of the progenitor stars for varied
supernova types remains an open subject. 

A supernova explosion from the core
collapse of a massive star drives a powerful shock wave into the
circumstellar medium (CSM) of the progenitor. The shock wave with a 
speed approximately 1000 times larger than the speed of the progenitor
star's wind quickly probes the circumstellar medium established
by the wind lost over many thousands of years before the explosion.
Relativistic electrons and magnetic fields in the interaction region
give rise to nonthermal radio emission. The interaction of a 
supernova shock with its surrounding medium produces X-ray emission
probing the environment. Multiwaveband
studies from radio and X-ray bands thus provide a handle on the
past history of the parent star and the nature of the interaction region.
Fingerprinting the exploded ejecta composition 
through X-ray spectroscopy \cite{Zim03} can give clues to the mass
of the exploding star (e.g. in SN 1998S \cite{Poo02}).

In this paper, we discuss several supernovae:
SN 2002ap -- a `hypernova' of type Ic 
associated with GRBs {\it that had no GRB counterpart}; another 
type Ic `hypernova'
SN 2003dh which was associated with GRB030329; and several other supernovae
such as: SN 2001gd, SN 2001ig and SN 2002hh. We also
mention how upcoming X-ray
spectroscopic studies of SN 1995N will help discriminate between
the sites of observed X-ray emission.

\section{`Hypernova' SN 2002ap: nature of its X-ray emission}
\label{sec:sn2002ap}

SN 2002ap showed early (`prompt') X-ray and radio emission. 
It was a nearby ($7.3 \ \rm Mpc$) and optically bright 
($V= 14.5$ upon discovery by Y. Hirose as in 
\cite{Nak02}) supernova. 
The broad spectral features in the optical (thus the
name `hypernova') and a subsequent modeling of its spectroscopic and
photometric data (\cite{Maz02}) suggested that it was energetic: 
$E_{\rm expl} \sim 4 - 10 \times 10^{51} \ \rm erg$.


XMM-Newton observed SN 2002ap with the EPIC
X-ray cameras and the Optical
Monitor for $\sim 34 \ ks$
on Feb 2.0 -- 2.4, 2002 UT, (day five after estimated 
explosion date Jan 28.0, 2002). A presupernova exposure of the field
by Chandra X-ray Observatory  on Oct. 19, 2001
revealed the presence of a nearby source about $14''.9$ away from 
the supernova, and the contribution of this source was taken into
account in our measurement of the day 5 SN X-ray flux:
$1.07^{+0.63}_{-0.31} \times 10^{-14} \ \rm erg \ cm^{-2} \ s^{-1}$
 ($0.3 - 10 \ \rm keV$)
(see \cite{Sut03} for details). Because
the SN was very faint, both thermal bremsstrahlung model ($N_H = 4.9 \times
10^{21} \ \rm cm^{-2}$, $kT = 0.8 \ \rm keV$) and a simple power law
model ($N_H = 4.2 \times 10^{21} \ \rm cm^{-2}$, photon index $\alpha = 2.6$)
fit the sparse data equally well. 

The earliest radio detection of SN 2002ap was $~ 4.5$ days after the
explosion, in the VLA 8.46 MHz band, and the frequency of the peak radio
flux declined from 8.46 GHz to 1.43 GHz over a period of 10 days from
the explosion date \cite{Ber02}. SN 2002ap was observed with the GMRT
at 610 MHz 8.96 days after explosion 
and yielded $2\sigma$ upper limit of $0.34 \ \rm mJy$ 
on Feb 5, 2002 \cite{Sut03}. The wavelength dependence of the radio
turn-on shifting to longer wavelengths 
can be due to either free-free absorption or  
synchrotron self-absorption in the expanding circumstellar matter
overlying the interaction region.
We have fitted the VLA and GMRT data on 2002
Feb. 5.96 (day 8.96) to a SSA model, with photon index $\alpha = -0.8$
in the optically thin limit, 
implying that the radius of the radio
photosphere on this day was $R_{\rm r} = 3.5 \times 10^{15} \ \rm cm$
and the magnetic field in the shocked ejecta was $B = 0.29 \ \rm G$. 
The SSA prediction of flux at 610 MHz band is 
consistent with the GMRT upper limit. 

\begin{figure}
\includegraphics[height=5.8cm]{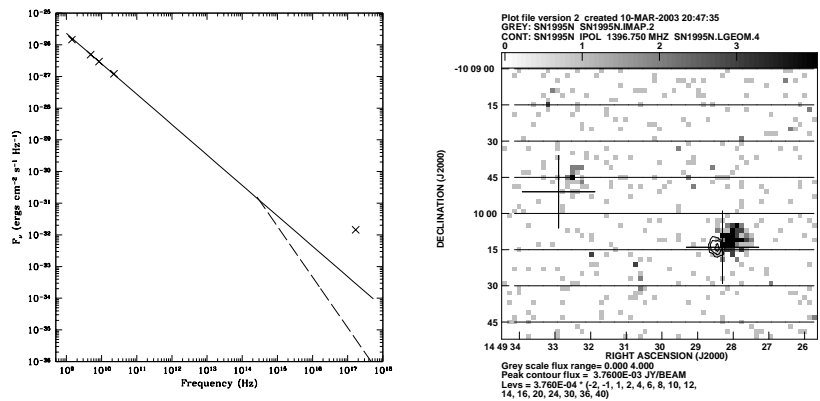}
\includegraphics[height=5.3cm]{alakray_fig2.eps}
\small\caption{Left: Multiwave plot for 2002ap. Dashed line shows the 
cooling break. Right: SN1995N GMRT 1420MHz contour labels 
overlayed on ROSAT HRI grey
image.
\label{fig:1}}
\end{figure}

The measured flux density of the SN 2002ap on day 5 corrected
for absorption: $F_{\nu}$ versus $\nu$ from the radio bands to the X-ray 
band is shown in Fig \ref{fig:1}. A single power-law
with the photon index $\alpha = -0.9$ (from the radio
spectrum) implies a flux density of only 58 picoJansky at 1 keV
(corresponding to $5\times 10^{-16} \ \rm erg s^{-1} cm^{-2}$ over
the effective bandwidth)
and fails to reproduce the observed X-ray flux.
A synchrotron radiation spectrum from a single popoulation
of relativistic electrons produces a power law spectrum with 
a steepening beyond the cooling
frequency. For the relevant parameters determined from radio
frequency spectral fits, this break should occur in
the optical region. A spectrum with such a break 
(Fig \ref{fig:1})
makes the observed radio and X-ray flux densities even more discrepant
than that with a constant $\alpha = -0.9$.

The observed X-ray flux by XMM-Newton could have been accounted for by
the thermal free-free emission (Bremsstrahlung). However, with the 
reported high ejecta velocity ( $v \geq 20,500 \ \rm km s^{-1}$ on
day 3.5) the implied temperature of the shocked ejecta and the
circumstellar matter a flat tail of high energy photons
upto about 100 keV  would have resulted \cite{Fra82} (only a limited cool
absorbing shell may have been present at this stage). By contrast, the
observed X-ray emission is quite soft 
(e.g. thermal Bremsstrahlung temperature $T_B = 0.8 \ \rm keV$). 
Even the reverse shock produced X-ray emission would be quite
hard compared with observed colours
unless the density gradient of the ejecta is extremely steep, -- 
normally not found for relevant progenitors.

The most natural explanation of the observed X-ray flux and spectral
features is the (repeated) Compton scattering of hot electrons
off optical photons from the photosphere at $T_{eff} \geq 10^4 \ K$. 
Detailed Monte Carlo simulations of the repeated Compton scatterings
have been performed by \cite{Poz77}.
The Compton flux is approximately
related to the optical flux by \cite{Fra82},\cite{Che94}:

$$ {\cal F}_{\nu}^{Compton} \sim \tau_{e} {\cal F}_{\nu}^{opt}
(\nu_o/\nu)^{\gamma} \rm erg \; s^{-1} cm^{-2} Hz^{-1} $$
   where the optical depth and the energy index are:
$$\tau_e  = {\dot M \sigma_T \over 4 \pi m_p R_s u_w} \Big(1 - {R_{opt} \over R_s}\Big) ; \; \;
\gamma (\gamma + 3) = -{m_ec^2\over kT_e} \rm ln\Bigl[{\tau_e\over
2}(0.9228- \rm ln \tau_e)
\Bigr] $$

We note that on day 5 the unabsorbed X-ray and optical flux densities
derived from XMM and ground based observations imply a multiwaveband
power-law index $\gamma = 2.5 - 2.8$ (somewhat steeper than the
XMM-band index $\gamma_{\rm XMM} \sim 1.6$) and a logarithm of the ratio
of flux densities of $\approx 7.4$. Typical electron temperature required
of Comptonizing plasma is $T_e = 1.5 - 2 \times 10^9 \ \rm K$ for
optical depths in the range $\tau_e = 4 - 25 \times 10^{-4}$ for
progenitor scenarios such as Wolf-Rayet stars or interacting binaries
involving Roche lobe overflow from a helium star. Such temperatures are
well wihin the range of hot circumstellar gas even for the modest velocities
of $16,000 - 20,000 \ \rm km s^{-1}$.

\section{SN 1995N, SN 2003dh and other supernovae}
\label{sec:grb+sne}

SN 1995N was a type IIn supernova that exploded in the galaxy MCG-2-38-017.
We have detected SN 1995N a number of times with the GMRT. Its
low frequency radio flux is very slowly declining (1420 MHz band
flux: $4.5 \pm 0.75 \ \rm mJy$ on 8 Nov 2000 and $4.2 \pm 0.2 \ \rm mJy$ 
on 21 Sep 2002; in 610 MHz band $3.3 \pm 0.35 \ \rm mJy$ on 16 Sep 2002).
Of the 19 supernovae \cite{Imm02}
that have been detected in the X-ray bands
SN1995N appears at the high end of the X-ray luminosity ($ \sim 1 \times 10^{41}
\ \rm erg s^{-1}$ \cite{Fox00}).
There is indication of short term variations of its X-ray
luminosity in the Rosat and ASCA observations. The Rosat HRI
X-ray image is overlaid with the GMRT 1400 MHz band
radio map of the region around SN 1995N in Fig. \ref{fig:1}. Here
the cross on the lower right is the position of the supernova while
that on the left is the galactic centre. The question of 
variation of its X-ray flux
will be addressed by our Chandra/XMM observations in the current cycle.
X-ray line-widths can also distinguish between the models of emission
by $1)$ ejecta gas struck by the reverse shock \cite{Che94} or $2)$ 
the radiative cooling of shocked dense clouds crushed by the strongly
shocked circumstellar wind \cite{Chu93}. Nucleosynthetic fingerprinting
through X-ray spectra will assist in determining
the progenitor star mass. 

\begin{figure}
\includegraphics[height=6.0cm]{alakray_fig3.eps}
\includegraphics[height=6.0cm]{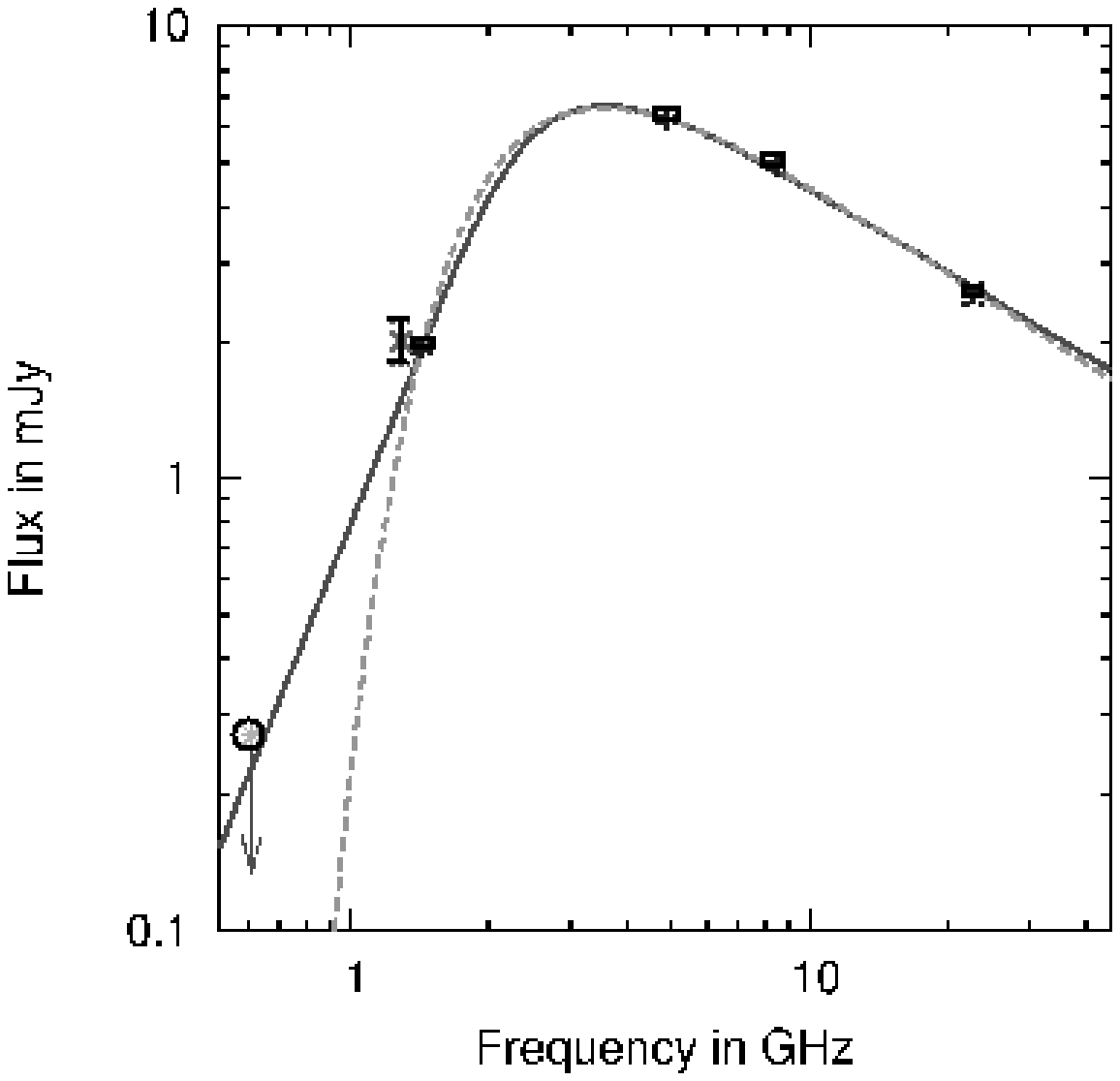}
\small\caption{Left: GMRT 1280MHz image of 2003dh \& Right: 
SSA (solid) and FF (dash) fits to the VLA (rectangle)\& GMRT (1280MHz: cross, 
610 upper limit: circle) 
\label{fig:2}}
\end{figure}

In contrast, SN 2003dh was an energetic type Ic supernova 
(a hypernova, with an
large expansion velocity 
$\sim 36,000 \pm 3000 \ \rm km \ s^{-1}$ and
estimated total isotropic energy release $\sim 9 \times 10^{51} \ \rm erg$)
that was spatially and temporally coincident with a GRB030329
at a redshift of $z=0.1685$.
We observed SN 2003dh with the GMRT in the 1280 MHz band and
on June 17, 2003 in the 
610 MHz band. The image of the region containing the SN is shown
in Fig. \ref{fig:2}. The flux of the SN was  $2.1 \pm 0.13\ \rm mJy$ in
the 1280 MHz band on 13 Jun 2003. The SN was not detected in the 610 MHz band.
We have combined our data with VLA measurements reported by Berger et al
\cite{Ber03} (June 4.01, 2003);
the resultant spectrum is hown in Fig. \ref{fig:2}.
The spectrum is consistent with both free-free and synchrotron
self-absorption (SSA) models. For the SSA model, the best fit parameters
are: $R=2.26 \times 10^{17} \ \rm cm$ and $B = 0.13 \ \rm G$. 
This is a relatively
high field for the radio emission region since for other
supernovae like SN 1998bw fields such as these are encountered
much earlier.

In the GMRT 1420 MHz band observations, SN 2001gd (IIb)
brightened marginally from
$3.2\pm0.3 \ \rm mJy (22Sep2002)$ to 
$3.6\pm0.2 \ \rm mJy (15Nov2002)$  while SN 2001ig (also type IIb)
dimmed between 
$14.5\pm0.4 \ \rm mJy (25Sep2002)$ to 
$9.5\pm0.3 \ \rm mJy (14Nov2002)$.
The 1400 MHz flux of SN 2002hh (type II)
on 27Dec2002 was $1.4\pm0.24 \ \rm mJy$.
We thank the staff of XMM-Newton and GMRT (NCRA-TIFR) that made these 
observations possible.


%
\index{paragraph}
%
%
%
%
%
%
%
%
%
%
%
%
%

%
%



\printindex
\end{document}